\documentclass[letterpaper,titlepage,11pt]{article}
\usepackage[usenames,dvipsnames]{xcolor}

\usepackage{amssymb,amsmath,amsfonts}

\usepackage[colorlinks,hyperindex, 
            bookmarks=true,bookmarksopen=true]{hyperref}
    \hypersetup
    {
        colorlinks=true,
        linkcolor=MidnightBlue,
        urlcolor=MidnightBlue,
        filecolor=black,
        citecolor=RedOrange,
        pdfstartview=FitV,
        pdftitle={},
        pdfauthor={Ilmar Gahramanov, Kemal Tezgin},
        pdfsubject={},
        pdfkeywords={},
        pdfpagemode=None,
        bookmarksopen=true
    }

\usepackage{mathtext}
\usepackage{stackrel}

\usepackage[utf8]{inputenc}

\begin{document} 

$\;$\\ 
\vspace{2.7cm}

\centerline{\Large \bf Remark on the Dunne-\"{U}nsal relation in exact semi-classics}
\vskip 1.3 cm

\centerline{\large {\bf Ilmar Gahramanov$^{a,b,c}$ and Kemal Tezgin$^{d}$}  }

{\small
\begin{center}
\textit{$^a$Max Planck Institute for Gravitational Physics (Albert Einstein Institute),\\ Am M\"{u}hlenberg 1, D-14476 Potsdam, Germany} \\
\vspace{2.5mm}
\textit{ $^b$Institute of Radiation Problems ANAS,\\ B.Vahabzade 9, AZ1143 Baku, Azerbaijan} \\
\vspace{2.5mm}
\textit{$^c$School of Engineering and Applied Science, Khazar University, \\ Mehseti Str. 41, AZ1096, Baku, Azerbaijan} \\
\vspace{2.5mm}
\textit{$^d$Department of Physics, University of Connecticut, \\ Storrs, CT 06269, USA} \\
\texttt{} \\
\vspace{.1mm}
\vspace{.1mm}
\href{mailto:ilmar@aei.mpg.de}{{\textbf{ilmar.gahramanov@aei.mpg.de}}} $\;\;\;\;\;\;\;\;$  \href{mailto:kemal.tezgin@uconn.edu}{{\textbf{kemal.tezgin@uconn.edu}}}
\end{center}
}

\vskip 0.5cm \centerline{\bf Abstract} \vskip 0.2cm \noindent  Recently, it is realized that non-perturbative instanton effects can be generated to all orders by perturbation theory around a degenerate minima via Dunne-\"{U}nsal relation in several quantum mechanical systems. In this work we verify the Dunne-\"{U}nsal relation for resonance energy levels of one-dimensional polynomial anharmonic oscillators. We show that the relation is applicable to cubic and quartic anharmonic oscillators which are genus one potentials. However for higher order (higher genus) anharmonic potentials the relation is not satisfied and is subject to a certain extension.\\ \vspace{0.4cm}

\line(2,0){140}

\newpage

\section{Introduction}

It has been well known that in quantum mechanical systems perturbation theory around a degenerate minima has energy expansion of the form
\begin{equation} \label{energy}
E(g)=\sum_{n=0}^\infty  E_n \, g^n,
\end{equation}
which often diverges asymptotically \cite{Brezin:1977gk,Stone:1977au,ZinnJustin:1989mi}. One way to resolve this issue is to apply Borel summation and give a physical meaning to these divergent series. However, in the process of analytic continuation of the Borel transform, singularities arise on the integration contour and hence the Borel sum includes imaginary terms due to the deformation of the contour (see, e.g., \cite{ZinnJustin:1981dx,ZinnJustin:1982td,ZinnJustin:1983nr,Aoyama:1998nt,Aoyama:1997qk,Unsal:2012zj,Aniceto:2013fka}). Moreover, the choice of the contour also affects the imaginary contribution hence gives rise to ambiguities on the energy eigenvalue \cite{Cherman:2014ofa, Aniceto:2013fka}. Another ambiguity arise from the fluctuations around the n-instanton sector which again asymptotically diverges and leads to ambiguous imaginary terms. So by the inclusion of these non-perturbative effects, we confront ambiguous imaginary terms coming from both perturbative and non-perturbative sectors which in the first sense make the problem even more subtle. Since any physical observable must be real and ambiguity free, a further analysis is needed to resolve these issues.

Recently, an important progress has been made in studying the question of the relation between the pertubative and non-perturbative contributions in quantum theories \cite{ZinnJustin:2004ib,ZinnJustin:2004cg,Jentschura:2004jg,Jentschura:2010zza, jentschura2009unified, Jentschura:2011zza} (see also earlier works \cite{Bender:1969si,Bender:1990pd,Lipatov:1976ny,ZinnJustin:1989mi,LeGuillou:1990nq}). Many of this progress is due to resurgence theory, developed by Ecalle in the early 1980 \cite{Ecalle} (see also \cite{Candelpergher,Delabaere}). Rather than the usual perturbative expansion (\ref{energy}) for the energy eigenvalue, resurgence analysis connects perturbative contributions with non-perturbative effects through ``resurgent trans-series'' that the imaginary terms with ambiguities coming from the Borel summation cancels systematically each other to all orders\footnote{For an introduction to resurgence in physics, see recent reviews on the topic \cite{Caliceti:2007ra,Shifman:2014fra,Dorigoni:2014hea}.} \cite{Dunne:2013ada,Dunne:2014bca}. For instance, imaginary term arising from the perturbative vacuum cancels the imaginary term arising from the 2-instanton sector, imaginary term from the 1-instanton sector is cancelled by an imaginary term in the 3-instanton sector and so on. Hence leaving us a real and unambiguous result for our observable (in this paper, energy). This cancellation has been carried to all orders by using the following resurgent expansion form of the $N$-th energy level \cite{ZinnJustin:2004ib,ZinnJustin:2004cg,Dunne:2013ada,Dunne:2014bca}
\begin{equation} \label{fullexpansion}
E^{(N)}(g)=\sum_{\pm}\sum_{n=0}^\infty \sum_{l=1}^{n-1}\sum_{m=0}^\infty c^{\pm}_{n, l, m}\, \frac{e^{-n\frac{S}{g}}}{g^{n(N+\frac{1}{2})}}\hskip -.1 cm  \left(\ln \left[\mp\frac{2}{g}\right]\right)^l \hskip -.1 cm g^{m}
\end{equation}
which takes into account the $n$-instan\-ton contributions, generated by $e^{-\frac{S}{g}}$ where $S$ being the coefficient of the instanton action, with the fluctuations around them as well as the quasi-zero-modes generated by the logarithmic term. One should note that for $n=0$ the term in the sum is the usual perturbative expansion of the form (\ref{energy}) and the logarithmic terms starts appearing at the $2$-instan\-ton sector. So the expansion (\ref{energy}) is actually not the complete expansion of a real unambiguous eigenvalue, one needs to extend it by adding non-perturbative effects. On the other hand, the expansion (\ref{fullexpansion}) handles these non-perturbative effects with the right coefficients $c^{\pm}_{n, l, m}$ such that the total sum is real and ambiguity free. This cancellations imply a deep relationship between perturbative and non-perturbative sectors and this relationship lies behind the resurgence analysis \cite{ZinnJustin:2004ib,ZinnJustin:2004cg,Dunne:2014bca}.

In recent years resurgence techniques have also been applied to different branches of physics and mathematics, including quantum mechanics \cite{ZinnJustin:2004ib,ZinnJustin:2004cg,Jentschura:2010zza,Dunne:2013ada,Dunne:2014bca,Krefl:2013bsa,Misumi:2015dua,Behtash:2015loa}, quantum field theory \cite{Argyres:2012ka,Dunne:2012ae,Dunne:2012zk,Cherman:2013yfa,Cherman:2014ofa,Dunne:2015ywa, Klaczynski:2016mbx}, string theory \cite{Marino:2007te, Marino:2008ya, Pasquetti:2009jg, Aniceto:2011nu, Marino:2012zq, Santamaria:2013rua,Couso-Santamaria:2014iia,Grassi:2014cla}, hydrodynamic gradient expansion \cite{Heller:2015dha,Aniceto:2015mto,Basar:2015ava}, supersymmetric theories \cite{Aniceto:2014hoa,Cherman:2014xia,Basar:2015xna,Behtash:2015kva}.

The organization of the paper is the following. In order to put the result of the notes in context, in Section 2 we briefly discuss some aspects of  \cite{Dunne:2013ada} and \cite{Jentschura:2010zza}, focusing in particular on the Dunne-\"{U}nsal relation. In Sections 3 and 4 we verify the relation for cubic and quartic anharmonic oscillators, respectively. While, the present work is mainly devoted to the verification of the Dunne-\"{U}nsal relation, we also briefly recall some properties of the cubic and quartic potentials. In Section 5 we discuss that the formula is not satisfied in its current form for higher degree polynomial oscillators, in particular we show the calculations for the quintic potential. For the sake of completeness, in Appendices A,B and C we write down the related data for sextic, septic and octic potentials, although they will not be discussed in the paper. In this paper the generalized quantization conditions and expressions for the functions $B$ and $A$  in terms of $E$ and $g$ are collected from the paper \cite{Jentschura:2010zza}.

\section{Connecting perturbative and non-perturbative sectors}

In Zinn-Justin et al. the resurgent expansion (\ref{fullexpansion}) can be obtained by a systematic small $g$ expansion of exact quantization condition \cite{ZinnJustin:1989mi,ZinnJustin:2004ib,ZinnJustin:2004cg,Jentschura:2010zza} in a given system. In their approach, this quantization condition includes two functions $B(E,g)$ and $A(E,g)$ which are related to the perturbative expansion of the energy eigenvalues and instanton contributions to the system, respectively \cite{ZinnJustin:1982td,ZinnJustin:1983nr}. Schematically, this generalized quantization condition has the following form
\begin{equation} \label{Zqc}
\frac{1}{\Gamma\left(1-B(E,g)\right)} \; \sim \;  \left(\frac{2}{g} \right)^{B(E,g)} \ e^{-A(E,g)} \;.
\end{equation}
One can compute the perturbative expansions of $A(E,g)$ and $B(E,g)$ functions by using the WKB approximation \cite{ZinnJustin:1983nr,ZinnJustin:2004ib,ZinnJustin:2004cg}. Alternatively, if one knows the energy in the one-instanton approximation to all orders, then from the one-instanton approximation of the quantization condition\footnote{Note that in case of locally harmonic oscillators, taking $B(E,g)=\frac12+N$ gives the usual perturbative expansion of the energy.} (\ref{Zqc}) it is easy to determine the function $A(E,g)$ \cite{ZinnJustin:1982td}. 

In order to calculate exact energy eigenvalues of a quantum-mechanical system as in the form of (\ref{fullexpansion}), one has to calculate the $B(E,g)$ and $A(E,g)$ functions separately and then expand this quantization condition for a small coupling parameter $g$. 

Lately, Dunne and \"{U}nsal have rather revealed a remarkable simple relation between these two functions. For given $B(E,g)$ and $A(E,g)$ functions of several physical systems (like double-well, sine-Gordon, Fokker-Planck, $O(d)$ symmetric potential) it was shown that by converting the functions $B(E,g)$ and $A(E,g)$ into $E(B,g)$ and $A(B,g)$ they satisfy the following relation \cite{Dunne:2013ada}
\begin{equation} \label{DunneUnsal}
\frac{\partial E(B, g)}{\partial B} =  -\frac{g}{2S} \left( 2B+g \frac{\partial A(B,g)}{\partial g} \right)
\end{equation}
where $S$ is the instanton action coefficient. 

The relation (\ref{DunneUnsal}) provides us a powerful computational tool: By knowing the perturbative expansion about a degenerate minima with a global boundary condition, one can derive the function $A$ rather than calculating them seperately. In other words, resurgent transseries for the energy can actually be generated only from the perturbative expansion of $E$ with a global boundary condition. The Dunne-\"{U}nsal relation (\ref{DunneUnsal}) shows us a close connection between perturbative and non-perturbative sectors which is not very obvious in the Zinn-Justin et al. approach \cite{ZinnJustin:1989mi,ZinnJustin:2004ib,ZinnJustin:2004cg,Jentschura:2010zza}. 

Although the Dunne-\"{U}nsal relation (\ref{DunneUnsal}) is a powerful equation, under which conditions this equation holds is still unclear \cite{Gorsky:2014lia,Dunne:2015eaa}. 

In this paper, we consider a set of one-dimensional anharmonic oscillators with polynomial potentials \cite{Jentschura:2004jg,Bender:1990pd,Bender:1969si,ZinnJustin:2010ng,Jentschura:2009jd} and verify the Dunne-\"{U}nsal relation for cubic and quartic anharmonic potentials which correspond to genus one potentials. However, as we go into higher order (higher genus) anharmonic potentials we observe that the relation is not satisfied. 

We use the notations of \cite{Jentschura:2010zza} and denote the Hamiltonian of an even oscillator by 
\begin{equation}
H_N(g) \ = \ -\frac12 \frac{\partial^2}{\partial q^2} + \frac12 q^2 + g q^N
\end{equation}
and  we use the convention  $H_M$ for the Hamiltonian of an odd oscillator\footnote{There are several reasons for choosing the coupling constant as $\sqrt{g}$ for odd oscillators. We will not stress these aspects in this paper, referring the reader to \cite{Jentschura:2010zza} for complete details.}
\begin{equation}
H_M(g) \ = \ -\frac12 \frac{\partial^2}{\partial q^2} + \frac12 q^2 + \sqrt{g} q^M.
\end{equation}
For $N$ even, instanton configurations exist for $g<0$ and the generalized Bohr-Sommerfeld quantization condition for each potential has the following form
\begin{align} \label{quant1}
& \frac{1}{\Gamma\left( \frac12 - B_N(E,g)\right)} =
\frac{1}{\sqrt{2 \pi}} \left(-\frac{ 2^{\frac{N}{N-2}}} {(-g)^{2/(N-2)}} \right)^{B_N(E,g)}  e^{  -A_N(E,g) } \;.
\end{align}
For $M$ odd, instantons exist for $g>0$ and the quantization condition reads
\begin{align} \label{quant2}
& \frac{1}{\Gamma\left( \frac12 - B_M(E, g)\right)} =
\frac{1}{\sqrt{8 \pi}} \left( - \frac{ 2^{\frac{M}{M-2}}} {g^{1/(M-2)}} \right)^{B_M(E, g)} e^{-A_M(E, g)} \;.
\end{align}
Here $\Gamma(z)$ is the Euler gamma function.

In \cite{Jentschura:2010zza, jentschura2009unified, Jentschura:2011zza}, Zinn-Justin et al. discuss contributions of instanton related effects in one-dimensional anharmonic oscillators of arbitrary even and odd degree, in particular they present expressions for $B(E, g)$ and $A(E, g)$ for the anharmonic  oscillators with polynomial potentials of degree $M=3,5,7$ and $N=4,6,8$. In this work we check the Dunne-\"{U}nsal relation for those potentials by using their generalized quantization conditions.

\section{Harmonic oscillator with cubic potential}

In this section we consider anharmonic oscillator with cubic potential. The Hamiltonian of cubic anharmonic oscillator has the following form
\begin{eqnarray}\label{cubH}
H_3(g) \ = \ -\frac12 \frac{\partial^2}{\partial q^2} + \frac12 q^2 + \sqrt{g} q^3 \;.
\end{eqnarray}
Note that for positive and real coupling parameter $g$ the one-dimensional cubic oscillator possesses resonances \cite{Jentschura:2007py, Jentschura:2010zza}. 

First we calculate the instanton action of the system (\ref{cubH}). The instanton action for odd anharmonic oscillators can be computed by the following general formula \cite{Jentschura:2009jd,Jentschura:2010zza}
\begin{equation} \label{oddinstanton}
s_M[q]=\Big(\frac{1}{g}\Big)^{1/(M-2)} \int_{0}^{2^{\frac{1}{2-M}}}2\sqrt{q^2-2q^M}dq \;,
\end{equation}
where $M$ stands for the degree of the polynomial. In case of the cubic potential, i.e. for $M=3$ we have
\begin{equation}
s_3=\frac{1}{g} \int_{0}^{\frac12}2q\sqrt{1-2q}dq \;.
\end{equation}
By defining $u=1-2q$, we obtain 
\begin{equation}
s_3=\frac{1}{g} \int_{0}^{1}\frac{(1-u)}{2}\sqrt{u}du \;.
\end{equation}
Now one can easily compute the integral and get the final result 
\begin{equation} \label{inscubic}
s_3=\frac{2}{15g} \;,
\end{equation}
which is positive for $g>0$. This quantity determines the leading contribution to the ground-state energy of order $\exp(-\frac{2}{15g})$.
The generalized Bohr-Sommerfeld quantization condition for the cubic potential reads 
\begin{equation} \label{cubicquant}
\frac{1}{\Gamma\left( \frac12 - B_3(E,g)\right)} =
\frac{1}{\sqrt{8 \pi}}\, \left( - \frac{8}{g} \right)^{B_3(E,g)}  \;
e^{-A_3(E,g)} \;,
\end{equation}
with the following characteristic functions
\begin{align}
B_3(E,g) & = E+\sum_{i=1}^\infty g^i b_{i+1}(E)\\
A_3(E,g) & = \frac{2}{15g}+\sum_{i=1}^\infty g^i a_{i+1}(E)
\end{align}
where $b_i$ and $a_i$ are polynomials of degree $i$ in $E$ and the term $\frac{2}{15g}$ is the instanton action (\ref{inscubic}). The expression (\ref{cubicquant}) is a relation for the resonance energies of the anharmonic oscillator with the cubic potential for $g>0$.  It is worth mentioning here that for $g<0$ the right-hand side of the expression (\ref{cubicquant}) is equal to zero.

The expressions for $B(E,g)$ and $A(E,g)$ for the cubic anharmonic oscillator were calculated in \cite{Jentschura:2007py,Jentschura:2009jd,Jentschura:2010zza}. The expansion of the perturbative function $B_3(E,g)$ in $g$ is given by
\begin{align} \nonumber
B_{\rm 3}(E, g) & = E + g \left( \frac{7}{16} +
\frac{15}{4} \, E^2 \right)
+ g^2 \, \left( \frac{1365}{64} \, E + \frac{1155}{16} \, E^3 \right)
\\ \nonumber
& \qquad + \, g^3 \, \left( \frac{119119}{2048} +
\frac{285285}{256} \, E^2 + \frac{255255}{128} \, E^4 \right) \\ \nonumber
& \qquad + g^4 \, \left( \frac{156165009}{16384} \,E  +
\frac{121246125}{2048} \, E^3 +
\frac{66927861}{1024} \, E^5 \right) \\ \nonumber
& \qquad  + \, g^5 \, \Big( \frac{10775385621}{262144} +
\frac{67931778915}{65536} E^2  +  \frac{51869092275}{16384} \, E^4 \\ \label{Bcubic}
& \qquad \qquad \quad  +
\frac{9704539845}{4096} \, E^6 \Big) + \ldots \;,
\end{align}
and the non-perturbative function $A_{\rm 3}(E, g)$ has the following expansion
\begin{align} \nonumber
A_{\rm 3}(E, g) & = \frac{2}{15\,g} +
g\, \left( \frac{77}{32} + \frac{141}{8} \, E^2 \right) +
g^2\, \left( \frac{15911}{128} \, E + \frac{11947}{32} \, E^3 \right)
\\ \nonumber
& \qquad \quad \; + g^3\, \left( \frac{49415863}{122880} +
\frac{6724683}{1024}\, E^2 +
\frac{5481929}{512}\, E^4 \right) \\ \nonumber
& \quad \; \qquad + g^4\, \left( \frac{2072342055}{32768} \, E  +
\frac{44826677}{128}\, E^3 +
\frac{733569789}{2048}\, E^5 \right) \\  \nonumber
& \quad \; \qquad + g^5\, \Big( \frac{404096853629}{1310720} + \frac{1100811938289}{163840}\, E^2 +
\frac{307346388279}{16384}\, E^4\\
& \qquad \qquad \qquad +
\frac{134713909947}{10240}\, E^6 \Big)+ \ldots \,.
\end{align}
In order to verify the Dunne-\"{U}nsal relation, we need to rewrite expansions of the function $A$ and the energy $E$ in terms of $B$ and $g$. We use an ansatz $E(B,g)=B-\sum_{j=1}^\infty p_{j+1}(B)g^j$, where $p_{j+1}(B)$ are polynomials of degree $(j + 1)$ in $B$. By inserting the ansatz into (\ref{Bcubic}) and by comparison of coefficients in each order of $g$ we get
\begin{align} \nonumber
E_3(B,g) & = B-g \left(\frac{7}{16}+\frac{15}{4} B^2\right)-g^2 \left(\frac{1155}{64}B+\frac{705}{16} B^3\right)\\ \nonumber
& \quad - g^3 \left(\frac{101479}{2048}+\frac{209055}{256} B^2+\frac{115755 }{128} B^4 \right) \\ \nonumber
& \quad - g^4 \left(\frac{129443349}{16384} B+\frac{77300685 }{2048} B^3+\frac{23968161}{1024} B^5\right) \\ \nonumber
& \quad - g^5 \Big(\frac{2375536317}{65536} + \frac{26541790065}{32768} B^2 + \frac{3601649205}{2048} B^4 \\ \label{energycub}
& \qquad \qquad + \frac{1412410545}{2048} B^6 \Big)+\ldots \;.
\end{align}
One can verify this result for the nonalternating perturbation series for the ground state energy without instanton effects, i.e. by inserting $B_3 = \frac12$ in (\ref{energycub})
\begin{equation}
E_{\rm ground}^{(3)}(g) = \frac12 -\frac{11}{8} g -\frac{465}{32} g^2 - \frac{39709}{128} g^3 -
\frac{19250805}{2048} g^4 + \ldots \, .
\end{equation}
This is exactly the result obtained by the Rayliegh-Schr\"{o}dinger perturbation theory.
Similarly writing $A_3$ in terms of $B$ and $g$ will yield
\begin{align} \nonumber
A_{\rm 3}(B, g) & = \frac{2}{15g} +  g \left(\frac{77}{32}+\frac{141}{8} B^2 \right)   + g^2 \left( \frac{13937}{128} B+\frac{7717}{32} B^3 \right)  \\ \nonumber 
& \quad + g^3 \left(\frac{43147783}{122880}+ \frac{5153379}{1024} B^2+\frac{2663129}{512} B^4\right) \\ \nonumber
& \quad + g^4 \left( \frac{1769452671}{32768} B + \frac{240109947}{1024} B^3 + \frac{282482109}{2048} B^5 \right) \\ \nonumber
& \quad + g^5 \Big( \frac{724731745353}{2621440} + \frac{3555387349941}{655360} B^2 + \frac{359377601583}{32768} B^4 \\
& \qquad \qquad + \frac{168844301703}{40960} B^6 \Big)+\ldots \;.
\end{align}
After these conversions, the new series obey the following relation
   \begin{align}
   \frac{\partial E_{3}(B, g)}{\partial B} =  -\frac{15}{2}B g- \frac{15}{2} g^2  \frac{\partial A_{ 3}(B, g)}{\partial g} \;.
   \end{align}   
It means that the Dunne-\"{U}nsal relation \cite{Dunne:2013ada,Dunne:2014bca} is satisfied for the cubic potential in the following form\footnote{The $S$ stands for the coefficient of (\ref{inscubic}). }
\begin{equation} \label{DUcubicr}
\frac{\partial E(B, g)}{\partial B} =  -\frac{g}{S} \left(B+ g \frac{\partial A}{\partial g} \right) \;.
\end{equation}
From the relation (\ref{DUcubicr}) it is obvious that the non-perturbative function $A_{\rm 3}(E, g)$ could actually be determined by the perturbative function $B_{3}(E, g)$.

Note that the relationship between A and B functions in the quantization conditions provided by \cite{ZinnJustin:2004ib,ZinnJustin:2004cg} may differ from each other. In particular, the function $A$ in quantization conditions (\ref{quant1})-(\ref{quant2}) appears as $\exp(-A)$ rather than $\exp(-A/2)$ which is the case for double-well and sine-Gordon potentials. This is the reason why we have the factor $-g^2/S$ instead of $-g^2/2S$ in front of $ \frac{\partial A}{\partial g}$.

\section{Quartic potential case}

In this section we consider the anharmonic oscillator with quartic potential. The Hamiltonian for quartic potential is 
\begin{equation}
H_4(g) = -\frac{1}{2} \frac{\partial^2}{\partial q^2}  + \frac{1}{2}  q^2 +  g q^4.
\end{equation}
Note that for $g<0$ the system has resonances.

Instanton action for even anharmonic oscillators has a slight different formula than the expression (\ref{oddinstanton}) for odd potentials \cite{Jentschura:2009jd,Jentschura:2010zza}
\begin{equation}
S_N[q]=\Big(-\frac{1}{g}\Big)^{2/(N-2)} \int_{0}^{2^{\frac{1}{2-N}}}2\sqrt{q^2-2q^N}dq \;.
\end{equation}
Here the label $N$ stands for the degree of the polynomial. In case of quartic potential, i.e. for $N=4$ we have
\begin{equation}
S_4=-\frac{1}{g}\int_{0}^{\frac{1}{\sqrt{2}}}2q\sqrt{1-2q^2}dq \;.
\end{equation}
By defining $u=1-2q^2$, we get 
\begin{equation}
S_4=-\frac{1}{g} \int_{0}^{1}\frac{\sqrt{u}}{2}du \;.
\end{equation}
Then the instanton action is
\begin{equation} \label{quarticinst}
S_4=-\frac{1}{3g}.
\end{equation}

The generalized Bohr-Sommerfeld quantization condition in the case of the quartic potential reads \cite{Jentschura:2010zza}
\begin{equation}
\frac{1}{\Gamma\left( \frac12 - B_4(E,g)\right)} =
\frac{1}{\sqrt{2 \pi}} \left( \frac{4}{g} \right)^{B_4(E,g)} \; e^{-A_4(E,g)} \; ,
\end{equation}
with the following perturbative $B$ and non-perturbative $A$ functions 
\begin{align}
B_4(E,g) & = E+\sum_{j=1}^\infty g^j b_{j+1}(E) \;,\\
A_4(E,g) & = -\frac{1}{3g}+\sum_{j=1}^\infty g^j a_{j+1}(E) \;.
\end{align}
The coefficients $b_j$ and $a_j$ are odd or even polynomials in $E$ of degree $j$. The evaluation of $B_4$ and $A_4$ in terms of series in variables $E$ and $g$ has been described in \cite{Jentschura:2010zza}. The first five orders of $B$ and $A$ functions are given by
\begin{align} \nonumber
B_4(E, g) & = E - g \left( \frac{3}{8} +
\frac{3}{2} \, E^2 \right)
+ g^2 \, \left( \frac{85}{16} \, E + \frac{35}{4} \, E^3 \right) \\ \nonumber
& \quad -  g^3 \, \left( \frac{1995}{256} +
\frac{2625}{32} \, E^2 + \frac{1155}{16} \, E^4 \right) \\
& \quad + g^4 \, \left( \frac{400785}{1024} \,E +
\frac{165165}{128} \, E^3 +
\frac{45045}{64} \, E^5 \right)
+ \ldots \, ,
\end{align}
\begin{align} \nonumber
A_4(E,g) & = -\frac{1}{3\,g} -
g\, \left( \frac{67}{48} + \frac{17}{4} \, E^2 \right) +
g^2\, \left( \frac{671}{32} \, E + \frac{227}{8} \, E^3 \right)
\\ \nonumber
& \quad - g^3\, \left( \frac{372101}{9216} +
\frac{125333}{384}\, E^2 +
\frac{47431}{192}\, E^4 \right)
 \\
& \quad + g^4\, \left( \frac{3839943}{2048} \, E +
\frac{82315}{16}\, E^3 +
\frac{317629}{128}\, E^5 \right)
+ \ldots \;.
\end{align}
Note that the leading term of the non-perturbative function $A(E,g)$ contains the instanton action as given in (\ref{quarticinst}). 
In order to check the Dunne-\"{U}nsal relation we convert the series $B(E,g)$ into $E(B,g)$ as
\begin{align} \nonumber
E_4(B,g) \ & = \ B +g \left(\frac{3}{8}+\frac{3 }{2} B^2 \right)  -g^2 \left(\frac{67 }{16} B+\frac{17}{4}B^3 \right) \\ \nonumber
& \quad + g^3 \left(\frac{1539}{256}+\frac{1707}{32} B^2 +\frac{375}{16} B^4 \right)  \\ \label{quarE}
& \quad - g^4 \left(\frac{305141}{1024} B+\frac{89165}{128} B^3+ \frac{10689}{64} B^5 \right) + \ldots \;,
\end{align}
The alternating usual perturbation series for the ground state can be derived by taking $B_4= \frac12$ in (\ref{quarE}) 
\begin{equation}
E_{\rm ground}^{(4)} = \frac12 + \frac{3}{4} g - \frac{21}{8} g^2 +
\frac{333}{16} g^3 - \frac{30885}{128} g^4 + \ldots \;.
\end{equation}
which agrees with results in \cite{Bender:1969si,ZinnJustin:1989mi}.

By converting $A(E,g)$ into $A(B,g)$ we get
\begin{align} \nonumber
A_4(B,g) & =-\frac{1}{3g} -g\left(\frac{67}{48}+\frac{17 }{4} B^2 \right)+g^2 \left( \frac{569}{32} B+\frac{125}{8}B^3 \right) \\ \nonumber
& \quad - g^3 \left( \frac{305141}{9216}+ \frac{89165}{384} B^2 + \frac{17815}{192} B^4 \right)\\
& \quad  -g^4 \left(\frac{91745}{256}B + \frac{133505}{64} B^3 +\frac{3595}{2} B^5 \right) +  \ldots \;.
\end{align}
One can then see that these series satisfy the following equation
   \begin{eqnarray}
   \frac{\partial E_{\rm 4}(B, g)}{\partial B}&=& 3 B g+ 3 g^2  \frac{\partial A_{\rm 4}(B, g)}{\partial g}
   \end{eqnarray}
which is in the form of
\begin{equation} \label{quarticDU}
\frac{\partial E(B, g)}{\partial B} =  -\frac{g}{S} \left(B+ g \frac{\partial A}{\partial g} \right).
\end{equation}
Thus we verify the Dunne-\"{U}nsal relation for the quartic potential.

\section{A comment on higher degree potentials}

The problem arises when we consider higher power polynomial potentials, which correspond to ${\rm genus}>1$. So far only genus one potentials have been approved satisfying the Dunne-\"{U}nsal relation\footnote{This issue has been discussed by several authors \cite{Gorsky:2014lia,Dunne:2015eaa}. For instance, in \cite{Gorsky:2014lia} it was claimed  that for ${\rm genus}=1$ potentials the Dunne-\"{U}nsal relation coincides with the equation of motion in the Whitham dynamics.}. However in this section we observe that for higher genus case the Dunne-\"{U}nsal relation is not satisfied in its current form. 

As an example, let us consider the anharmonic oscillator with the polynomial potential of degree five (quintic). The Hamiltonian in this case is
\begin{equation}
H_5(g) \ = \ -\frac{1}{2} 
\frac{\partial^2}{\partial q^2}  + \frac{1}{2}  q^2 +  \sqrt{g}  q^5 \;,
\end{equation}
and the generalized quantization condition reads
\begin{equation}
\frac{1}{\Gamma\left( \frac12 - B_5(E,g)\right)} = \frac{1}{\sqrt{8 \pi}} \left( - \frac{ 2^{5/3} }{g^{1/3}} \right)^{B_5(E,g)} e^{-A_5(E,g)} \;.
\end{equation}
The instanton action for the ground state of the quintic potential can be calculated from the expression (\ref{oddinstanton}) and one gets the following result
\begin{align} \nonumber
s_5[q] & = \Big(\frac{1}{g}\Big)^\frac{1}{3} \int_{0}^{2^{-1/3}}2q\sqrt{1-2q^3}dq \\
& = \frac{3\,\sqrt{3}\,\,\Gamma^3(\tfrac{2}{3})}{7 \, \pi\,(2 g)^{1/3}} \;.
\end{align}

The perturbative function $B(E,g)$ has the following expansion
\begin{align} \nonumber
 B_5(E, g) & = E + g \left( \frac{1107}{256} +
\frac{1085}{32}  E^2 + \frac{315}{16} E^4 \right) \\ \nonumber
& \quad + g^2 \left( \frac{118165905}{8192}  E +
\frac{96201105}{2048}  E^3 +
\frac{15570555}{512}  E^5 +
\frac{692835}{128}  E^7 \right) \\ \nonumber
& \quad + \; g^3 \left( \frac{36358712597025}{4194304} + \frac{142306775756145}{1048576}  E^2 +
\frac{30926063193025}{131072} E^4 \right. \\ 
& \qquad \qquad \left. +
\frac{4140194663605}{32768}  E^6 +
\frac{456782651325}{16384}  E^8 +
\frac{9704539845}{4096} E^{10} \right)
+ \ldots \;.
\end{align}
By converting $E$ as a function of $B$ and $g$ for the first three terms we find that
\begin{align} \nonumber
E(B,g) & = B - g \left( \frac{1107}{256}+\frac{1085}{32} B^2+\frac{315}{16} B^4 \right)  \\ \nonumber
& \quad - g^2 \left(\frac{115763715}{8192}B +\frac{90794795}{2048} B^3 + \frac{13519905}{512} B^5+\frac{494385}{128} B^7 \right) \\ \nonumber
& \quad - g^3 \left(\frac{36099752507685}{4194304} + \frac{140162880546045}{1048576} B^2 + \frac{29646883011725}{131072} B^4 \right. \\
& \qquad \qquad \left. + \frac{3708489756265}{32768} B^6 + \frac{351124790625}{16384} B^8 + \frac{5590822545}{4096} B^{10} \right)
+\ldots \;.
\end{align}
The first few terms of the perturbative expansion of the function $A(E,g)$ for the quintic potential is given by \cite{Jentschura:2010zza}
\begin{align} \nonumber
A_5(E,g) & =  \frac{3 \sqrt{3} \Gamma^3(\tfrac{2}{3})}{7  \pi (2 g)^{1/3}} -
g^{1/3} \frac{3 \sqrt{3} \Gamma^3( \tfrac{1}{3} )}{ 2^{2/3}  8  \pi} 
\left( \frac{11}{54} + \frac{14}{27}  E^2 \right) \\ \nonumber
& \quad + g^{2/3}\, \frac{ \Gamma^3(\tfrac{2}{3} ) }{ 2^{1/3} \sqrt{3} \pi }
\left( \frac{385}{32} E + \frac{ 935 }{ 72 }  E^3 \right) \\ 
& \quad + g \left( \frac{21171}{1024}  + \frac{132245}{1152} E^2 + \frac{10865}{192} E^4 \right) + \ldots \;.
\end{align}
The function $A$ can be written in terms of variables $B$ and $g$ as follows
\begin{align} \nonumber
A_5(B,g) & =  \frac{3 \sqrt{3} \Gamma^3(\tfrac{2}{3})}{7  \pi (2 g)^{1/3}} -
g^{1/3}\, \frac{3 \sqrt{3} \,
\Gamma^3( \tfrac{1}{3} )}{ 2^{2/3} \, 8 \, \pi} \,
\left( \frac{11}{54} + \frac{14}{27} \, B^2 \right) \\ \nonumber
& \quad + g^{2/3}\, \frac{ \Gamma^3(\tfrac{2}{3} ) }{ 2^{1/3} \, \sqrt{3}\, \pi }\,
\left( \frac{385}{32} \, B + \frac{ 935 }{ 72 } \, B^3 \right) \\ 
& \quad + g\, \left( \frac{21171}{1024}  + \frac{132245}{1152}\, B^2 + \frac{10865}{192}\, B^4 \right) + \ldots \;.
\end{align}
One can easily see that this example differs from the preceding ones since the expansion includes Euler gamma functions and fractional orders of the coupling parameter $g$.

By using these data, the left hand side of the Dunne-\"{U}nsal relation yields 
\begin{align} \nonumber
\frac{\partial E}{\partial B} & = 1- g \left(\frac{1085}{16} B+\frac{315}{4} B^3 \right)  \\ \nonumber
& \quad - g^2 \left( \frac{115763715}{8192} +\frac{272384385}{2048} B^2 + \frac{67599525}{512} B^4+\frac{3460695}{128} B^6 \right) \\ \nonumber
& \quad - g^3 \left(\frac{140162880546045}{524288} B + \frac{29646883011725}{32768} B^3 \right. \\ \label{leftDU}
& \qquad \qquad \left. + \frac{11125469268795}{16384} B^5 + \frac{351124790625}{2048} B^7 + \frac{27954112725}{2048} B^{9} \right)
+\ldots \;
\end{align}
and from the right hand side we again get fractional powers of $g$ as well as gamma functions which clearly do not match to the left hand side (\ref{leftDU}) of the Dunne-\"{U}nsal relation. 

Other higher order (higher genus) potentials  have similar fractional terms in the expansion of $A(E,g)$ and they do not match with the left hand side of the Dunne-\"{U}nsal relation (\ref{DunneUnsal}). We provide $A$ and $B$ functions of these potentials for $N=6,8$ and $M=7$ in the appendices for convenience.

\section{Conclusions}

To conclude, trans-series expansion of an energy eigenvalue gives us a real and unambiguous result due to the cancellation of ambiguous imaginary terms arising from the perturbative expansion around perturbative vacuum and non-perturbative saddles. This cancellation mechanism implies a close relationship between perturbative and non-perturbative sectors. This cannot be easily seen in the Zinn-Justin et al. approach where one needs to separately calculate the functions $A(E,g)$ and $B(E,g)$. However, this relationship can be seen by the Dunne-\"{U}nsal relation \cite{Dunne:2013ada,Dunne:2014bca} and the non-perturbative sector can be generated purely from the perturbative sector. Rather than calculating the functions $A(E,g)$ and $B(E,g)$ seperately, it is actually enough to generate the trans-series expansion of energy by the knowledge of the perturbative function $B(E,g)$ with a global boundary condition. The Dunne-\"{U}nsal relation has been shown to apply for several genus one potentials including double-well, periodic sine-Gordon (periodic cosine), $O(d)$ symmetric and Fokker-Planck potentials.

In our current study we confirmed that the relation also holds for resonance energy levels of unified even and odd degree anharmonic complexified potentials. We verified the relation for cubic and quartic anharmonic potentials which are genus one potentials. However for higher order (higher genus) potentials we observed that the formula is not satisfied and needs to be generalized.

\vspace{0.3cm}

\noindent \textbf{Acknowledgments.} We express our sincere gratitude to Mithat \"{U}nsal for his careful reading of the manuscript, many useful comments and sharing his ideas. KT thanks to Gerald Dunne for many useful discussions, comments and suggestions.

\appendix
\section*{Appendices}

Here we list expansions of $A$ and $B$ functions in terms of series in variables $E$ and $g$, as well as generalized quantization conditions for sextic, septic and octic potentials. All the expressions for $B(E,g)$ and $A(E,g)$ listed in appendices A, B and C were taken from \cite{Jentschura:2010zza}.

\section{Sextic Potential}

The sextic anharmonic oscilator is described by the following Hamiltonian 
\begin{equation}
H_6(g) = -\frac{1}{2} \, \frac{\partial^2}{\partial q^2}  + \frac{1}{2} q^2 + g q^6 \;.
\end{equation}
The generalizaed Bohr-Sommerfeld quantization condition in this case reads
\begin{equation}
\frac{1}{\Gamma\left( \frac12 - B_6(E,g)\right)} =
\frac{1}{\sqrt{2 \pi}}
\left( \frac{2^{3/2}}{g^{1/2}} \right)^{B_6(E,g)} 
e^{-A_6(E,g)} \;.
\end{equation}
The first few terms of the perturbative expansions of the functions $A$ and $B$ for the sextic potential are given by
\begin{align} \nonumber
 B_6(E, g) & = E - g \left( \frac{25}{8} E +
\frac{5}{2} \, E^3 \right)  + g^2 \left( \frac{21777}{256} E + \frac{5145}{32} E^3
+ \frac{693}{16} E^5 \right) \\ 
& \quad - g^3 \left( \frac{12746305}{2048} E + \frac{8703695}{512} E^3
+ \frac{1096095 }{128} E^5 + \frac{ 36465 }{32} E^7 \right)
+ \ldots \;, \\ \nonumber
A_6(E,g) & = \frac{ \pi }{2^{5/2} (-g)^{1/2}} - g\left( \frac{221}{24}  E + \frac{17}{3} E^3 \right) \\ 
& \quad + g^2 \left( \frac{ 2504899 }{ 7680 } E + \frac{ 45769 }{ 96 }E^3 +
   \frac{ 17527 }{ 160 }E^5  \right)
+ \ldots \;,
\end{align}
where again the first term of the function $A$ has fractional order of the coupling parameter $g$. 

We also present here expansions of the energy $E$ and the non-perturbative function $A$ in terms of $B$ and $g$ for the sextic anharmonic oscillator:
\begin{align} \nonumber
E_6(B,g) & = B + g \left( \frac{25}{8} B + \frac{5}{2} B^3 \right)  - g^2 \left(\frac{19227}{256}B + \frac{4145}{32} B^3 + \frac{93}{32} B^5 \right)\\ \label{E6}
& \quad + g^3 \left( \frac{11719955}{2048} B + \frac{7364155}{512} B^3 + \frac{49245}{8} B^5 +\frac{28605}{32} B^7 \right) + \ldots \;, \\ \nonumber
A_6(B,g) & = \frac{ \pi }{2^{5/2} \, (-g)^{1/2}}
- g \left( \frac{221}{24} \, B + \frac{17}{3} \, B^3 \right) \\ 
& \quad + g^2  \left( \frac{1620899}{7680} B + \frac{23159}{96} B^3 + \frac{10727}{160}B^5 \right)
+ \ldots \;.
\end{align}
From these expressions one can easily see that the Dunne-\"{U}nsal formula is not satisfied. By taking $B=\frac12$ in (\ref{E6}) one obtains the following perturbative series for the ground state energy
\begin{equation}
E_{\rm ground}^{(6)}(g) = \frac12 + \frac{15}{8}g -\frac{3495}{64} g^2 +
\frac{1239675}{256}\, g^3 + \ldots \;.
\end{equation}

\section{Septic Potential}

The Hamiltonian of the septic anharmonic oscillator is
\begin{equation}
H_7(g) \ = \ -\frac{1}{2} \frac{\partial^2}{\partial q^2}  + \frac{1}{2} q^2 + \sqrt{g}  q^7 \;.
\end{equation}
The generalized quantization condition has the following form
\begin{equation}
\frac{1}{\Gamma\left( \frac12 - B_7(E,g)\right)} =
\frac{1}{\sqrt{8 \pi}}
\left( -\frac{2^{7/5}}{g^{1/5}} \right)^{B_7(E,g)} 
e^{-A_7(E,g)} \;.
\end{equation}
The first terms of the expansions for $B_7(E,g)$  and $A_7(E,g)$ read as follows
\begin{align} \nonumber
 B_7(E, g) & = E + g \left( \frac{180675}{2048} +
\frac{444381}{512}  E^2 + \frac{82005}{128} E^4 + \frac{3003}{32} E^6 \right) \\ \nonumber
& \quad + g^2 \, \left( \frac{182627818702875}{2097152} E +
\frac{ 156916927352185 }{  524288 } E^3 + \frac{ 13513312267455}{65536} E^5  \right. \\ & \qquad \left. + \frac{ 824707412529 }{ 16384 } E^7 + \frac{ 43689020375 }{ 8192 } E^9 +\frac{456326325}{2048}E^{11}  \right)
+ \ldots \;, \\ \nonumber
 A_7(E,g) & = \frac{5^{1/4} \, \Gamma(\tfrac{1}{5} )\,\Gamma(\tfrac{2}{5})}%
  {2^{1/10} \, ( \sqrt{5} + 1 )^{1/2}\,9 \, \pi \, g^{1/5}} +
g^{1/5} \frac{5^{1/4} \,
\Gamma^2( \tfrac{3}{5} )\,\Gamma( \tfrac{4}{5} )}%
{2^{9/10} \, (\sqrt{5} + 1)^{1/2}  \pi} \,
\left(\frac{5}{8} + \frac{9}{10}  E^2 \right) \\ \nonumber
 & \quad - g^{2/5}\, \frac{5^{1/4}  ( \sqrt{5} + 1 )^{1/2} 
\Gamma^2( \tfrac{1}{5} ) 
\Gamma( \tfrac{3}{5} ) }{2^{3/10}  \pi }
\left( \frac{377}{1600} E + \frac{299}{2000} E^3 \right) \\
& \quad + g^{3/5}  \frac{5^{1/4}  \Gamma( \tfrac{2}{5} )
\Gamma^2( \tfrac{4}{5} ) }{2^{7/10} ( \sqrt{5} - 1 )^{1/2}  \pi }
\left( \frac{59143}{9600} + \frac{15351}{400} E^2 + \frac{13209}{1000} E^4 \right)
+ \ldots \;.
\end{align}

The expressions of $E$ and $A$ in terms of $B$ and $g$ are then
\begin{align} \nonumber
E_7(B,g) & = B - g \left(\frac{180675}{2048} + \frac{444381}{512} B^2 + \frac{82005}{128} B^4 + \frac{3003}{32} B^6 \right) \\ \nonumber
& \quad - g^2 \left( \frac{182306664554175}{2097152} B + \frac{156008499432541}{  524288} B^3 + \frac{13291408081875}{65536} B^5 \right. \\ \label{E7}
& \qquad \qquad \left. + \frac{787132323285}{16384} B^7 + \frac{38763800075}{8192} B^9 + \frac{348110217}{2048} B^{11} \right)+\ldots \;, \\ \nonumber
A_7(B,g) & = \frac{5^{1/4} \Gamma(\tfrac{1}{5} ) \Gamma(\tfrac{2}{5})}%
  {2^{1/10} \, ( \sqrt{5} + 1 )^{1/2} 9  \pi  g^{1/5}} +
g^{1/5} \frac{5^{1/4} \Gamma^2( \tfrac{3}{5} ) \Gamma( \tfrac{4}{5} )}%
{2^{9/10}  (\sqrt{5} + 1)^{1/2} \pi} 
\left(\frac{5}{8} + \frac{9}{10} B^2 \right) \\ \nonumber
 & \quad - g^{2/5} \frac{5^{1/4} \, ( \sqrt{5} + 1 )^{1/2} \Gamma^2( \tfrac{1}{5} ) 
\Gamma( \tfrac{3}{5} ) }{2^{3/10}  \pi } \left( \frac{377}{1600} B + \frac{299}{2000}  B^3 \right) \\
& \quad + g^{3/5} \frac{5^{1/4}  \Gamma( \tfrac{2}{5} ) 
\Gamma^2( \tfrac{4}{5} ) }{2^{7/10}  ( \sqrt{5} - 1 )^{1/2} \pi } \left( \frac{59143}{9600} + \frac{15351}{400} B^2 + \frac{13209}{1000} B^4 \right)
+ \ldots \;.
\end{align}
Taking $B=\frac12$ in (\ref{E7}) gives rise to the following perturbative series for the ground state energy
\begin{align}
E_{\rm ground}^{(7)}(g) =& \frac12 - \frac{ 44379 }{ 128 } g
- \frac{ 715842493569  }{ 8192 }g^2 + \ldots \;.
\end{align}

\section{Octic Potential}

The Hamiltonian of the octic anharmonic oscillator is 
\begin{equation} 
H_8(g)  =  -\frac{1}{2} \frac{\partial^2}{\partial q^2}  + \frac{1}{2}  q^2 + g  q^8 \;.
\end{equation}
The generalized quantization condition is given by
\begin{equation}
\frac{1}{\Gamma\left( \frac12 - B_8(E,g)\right)} =
\frac{1}{\sqrt{2 \pi}}
\left( \frac{2^{4/3} }{g^{1/3}} \right)^{B_8(E,g)} 
e^{-A_8(E,g)} \;.
\end{equation}

The first few terms of the functions $B_8(E,g)$ and $A_8(E,g)$ are
\begin{align} \nonumber
B_8(E, g) & = E - g \left( \frac{315}{128} +
\frac{245}{16} \, E^2 + \frac{35}{8} \, E^4 \right) \\
& \quad + g^2 \, \left( \frac{5604849}{2048}  E + \frac{3209745}{512}  E^3
+ \frac{291291}{128} \, E^5 + \frac{6435}{32}  E^7 \right) + \ldots \;, \\ \nonumber
A_8(E,g) & = \frac{\sqrt{3}\, \Gamma^3( \tfrac{1}{3} ) }%
{2^{2/3}\,10\,\pi\, (-g)^{1/3}}
+ (-g)^{1/3} \, \frac{ \Gamma^3( \tfrac{2}{3} ) }%
{2^{1/3}\,\sqrt{3}\,\pi} \,
\left( \frac{17}{16} + \frac{5}{4}  E^2 \right) \\
& \quad - (-g)^{\frac23} \frac{ \Gamma^3( \tfrac{1}{3} ) }%
{2^{\frac23} \sqrt{3} \pi} \,
\left( \frac{77}{96} + \frac{91}{216} E^2 \right)
- g \left( \frac{28007}{2560} +
\frac{22669}{576} E^2 +
\frac{2587}{288} E^4 \right)
+ \ldots \;.
\end{align}

The expressions of $E$ and $A$ in terms of $B$ and $g$ then become
\begin{align} \nonumber
E(B,g) & = B + g \left(\frac{315}{128}+ \frac{245}{16} B^2 + \frac{35}{8} B^4 \right) \\ \label{E8} 
& \quad - g^2 \left(\frac{5450499}{2048} B + \frac{2947595}{512} B^3 + \frac{239841}{128} B^5 + \frac{3985}{32} B^7 \right)+ \ldots \;,
\\ \nonumber
A_8(B,g) & = \frac{\sqrt{3} \Gamma^3( \tfrac{1}{3} ) } {2^{2/3}\,10\,\pi\, (-g)^{1/3}} + (-g)^{1/3}  \frac{ \Gamma^3( \tfrac{2}{3} ) } {2^{1/3} \sqrt{3} \pi} 
\left( \frac{17}{16} + \frac{5}{4} B^2 \right) \\
& \quad - (-g)^{\frac23} \frac{ \Gamma^3( \tfrac{1}{3} ) }
{2^{\frac23} \sqrt{3} \pi} \left( \frac{77}{96} + \frac{91}{216} B^2 \right)
- g \left( \frac{28007}{2560} +\frac{22669}{576} B^2 + \frac{2587}{288} B^4 \right)
+ \ldots \;.
\end{align}

By taking $B=\frac12$ in (\ref{E8}) we find the following perturbative series for the ground state energy for the octic anharmonic oscillator
\begin{equation}
E_{\rm ground}^{(8)}(g) = \frac12 + \frac{105}{16} g - \frac{67515}{32} g^2 + \ldots \;.
\end{equation}


\end{document}